\documentclass[aps,10pt,final,
notitlepage, oneside, nobibnotes, %nofootinbib,
showpacs, centertags,
]{revtex4}

\newcommand\ba{\begin{eqnarray}}
\newcommand\ea{\end{eqnarray}}
\newcommand\be{\begin{equation}}
\newcommand\ee{\end{equation}}
\newcommand\nn{\nonumber}

\usepackage{hyperref}
\begin{document}
\def\ge{G_E(Q^2)}
\def\gm{G_M(Q^2)}
\def\mugegm{\mu_p G_E(Q^2) / G_M(Q^2)}
%\def\nn{\nonumber}

%%% article title
\title{Possible Method for Measuring the Proton Form Factors \\in Processes with and without Proton Spin Flip}

%%% author(s)
\author{M.~V.~Galynskii}
\email{galynski@dragon.bas-net.by}
\affiliation{ Joint Institute for Power and Nuclear Research-``Sosny'', National Academy of Sciences of Belarus, Minsk, 220109 Belarus}

\author{E.~A.~Kuraev}
\email{kuraev@theor.jinr.ru}
\affiliation{Joint Institute for Nuclear Research, Dubna, Russia}

\author{Yu.~M.~Bystritskiy}
\email{bystr@theor.jinr.ru}
\affiliation{Joint Institute for Nuclear Research, Dubna, Russia}

%%% abstract
\begin{abstract}
The ratio of the squares of the electric and magnetic proton form factors is shown to be
proportional to the ratio of the cross sections for the elastic scattering of an unpolarized
electron on a partially polarized proton with and without proton spin flip. The initial proton
at rest should be polarized along the direction of the motion of the final proton.
Similar results are valid for both radiative $ep$ scattering and the photoproduction
of pairs on a proton in the Bethe--Heitler kinematics. When the initial proton is fully
polarized in the direction of the motion of the final proton, the cross section for the
$ep \rightarrow ep$ process, as well as for the $ep \rightarrow ep \gamma$ and
$\gamma p \to e \bar e p$ processes, without (with) proton spin flip is
expressed only in terms of the square of the electric (magnetic) proton form factor.
Such an experiment on the measurement of the cross sections without and with proton
spin flip would make it possible to acquire new independent data on the behavior
of $G_E^2(Q^2)$ and $G_M^2(Q^2)$, which are necessary for resolving the contradictions
appearing after the experiment of the JLab collaboration on the measurement of
the proton form factors with the method of polarization transfer from the initial
electron to the final proton.
\end{abstract}

\pacs{13.40.Gp, 13.60.-r, 25.40.Ep}

\maketitle

\section*{Introduction}
% ==============================================================================
The investigation of the proton electromagnetic form factors, which are very important
characteristics of this fundamental object, provides a deeper insight into the structure
of the proton and the properties of the interaction between the constituent quarks.

Since the mid-1950s \cite{Rosen, Hof58}, to obtain the experimental data on the behavior of the proton
electric, $G_E(Q^2)$, and magnetic, $G_M(Q^2)$, form factors (Saks form factors) and to
analyze the electromagnetic structure of the proton, the electron--proton elastic
scattering has been used. For the case of the unpolarized electrons and protons,
all experimental data on the proton form factors were obtained with the Rosenbluth
formula \cite{Rosen} corresponding to the differential cross section of the elastic
$ep \to ep$ process,
\ba
\label{Ros}
\overline{\sigma}= \frac{d\sigma}{d\Omega}=\frac{\alpha^2E_e'\cos^2(\theta_e/2)}{4E_e^3\sin^4(\theta_e/2)}
\frac{1}{1+\tau} \left(\,G_E^2 +\frac{\tau}{\varepsilon}\,G_M ^2\right)\,.
\ea
Here, $\tau=Q^2/4M^2$, where $M$ is the proton mass and $Q^2=-q^2=4E_eE_e'\sin^2(\theta_e/2)$;
$E_e$ and $E_e'$ are the energies of the initial and final electrons, respectively;
$\theta_e$ is the electron scattering angle in the rest frame of the initial proton;
and the degree of the transverse (linear) polarization of a virtual photon,
$\varepsilon$, is determined by the expression $\varepsilon^{-1}=1+2(1+\tau)\tan^2(\theta_e/2)$.

According to the Rosenbluth formula, the leading contribution to the $ep \to ep$
cross section for high $Q^2$ values comes from the term proportional to the proton magnetic
form factor squared $G_M^2(Q^2)$; therefore, the accuracy of the separation of the $G_E^2(Q^2)$
contribution decreases. For this reason, the use of the Rosenbluth formula for experimentally
determining the form factors $\ge$ and $\gm$ gives significant uncertainties at
$Q^2\geq 1$ $GeV^2$.

Note that Rosenbluth formula (\ref{Ros}) valid in the laboratory reference frame, where the initial
proton is at rest, is naturally represented as the sum of two terms proportional to the
squares of the Saks form factors, $G^2_E$ andè $G^2_M$,
\ba
\label{Ross}
\overline{\sigma}=\sigma_{\uparrow\uparrow} + \sigma_{\uparrow\downarrow},\;\;
\sigma_{\uparrow\uparrow}=\kappa \, G^2_E ,\;\; \sigma_{\uparrow\downarrow}=\kappa \frac{\tau}{\varepsilon} \, G^2_M\;.
\ea
Here, $\kappa$ is the factor in front of the parentheses in Eq. (\ref{Ros}).
However, the physical meaning of these terms is not explained in the literature and is
unknown for most researchers. To elucidate the physical meaning of the terms
$\sigma_{\uparrow\uparrow}$ è $\sigma_{\uparrow\downarrow}$,
the following simple consideration is sufficient. The scattering cross section disregarding the
polarizations of the initial and final protons can always be represented as the sum of the cross
sections without and with the spin flip of the initial proton, which should be fully polarized
along a certain direction determined by the process kinematics. Since the initial proton is at
rest, this separated direction can be only the direction of the motion of the scattered proton.
Then, according to the commonly known additional reasonings (see, e.g., Eqs. (4.55) from \cite{RekBook}
or Eqs. (8.55) and (8.56) from \cite{Kallen}), in the Breit system of the initial and final protons,
the matrix element of the proton current for the case of the transition with change in helicity
(without spin flip) is expressed only in terms of the electric form factor $G_E$, whereas the
matrix element of the proton current for the case of the transition without change in helicity
(with spin flip) is expressed only in terms of the magnetic form factor $G_M$. Thus,
the terms $\sigma_{\uparrow\uparrow}$ and $\sigma_{\uparrow\downarrow}$ in Eq. (\ref{Ross})
 are the cross sections without and with the spin flip for the case where the initial proton
 is fully polarized in the direction of the motion of the final proton. Below, we demonstrate
 that our simple physical consideration is based on rigorous mathematical results obtained
 using the approach of the diagonal spin basis \cite{Sikach84,GS98}. Since the cross sections
 without and with the proton spin flip in Eq. (\ref{Ross}) are expressed only in terms of one of
 the Saks form factors, they can be attractable for performing direct experiments on their
 measurement and acquiring new independent data on the behavior of $G^2_E$ and $G^2_M$
 as functions of $Q^2$.

Akhiezer and Rekalo \cite{Rekalo} proposed a method for measuring the ratio of the Saks form factors
that is based on the polarization transfer from the longitudinally polarized initial electron
to the final proton and is independent of the Rosenbluth technique. In \cite{Rekalo}, it was shown that
the ratio of the degrees of the longitudinal, $P_l$, and transverse,
$P_t$, polarizations of the scattered proton is proportional to the ratio
of the proton electric and magnetic form factors:
\ba
\frac{P_l}{P_t}=-\frac{G_M}{G_E}\frac{E_e+E_e'}{2M} \tan\left ( \frac{\theta_e}{2}\right )\,.
\ea
The experiments based on the method of the polarization transfer from the initial electron
to the final proton were recently performed with high accuracy by the Bates \cite{Bates} and
JLAB \cite{PRL} Collaborations. They gave surprising results according to which $\ge$ decreases with an
increase in $Q^2$ faster than $\gm$) does; this contradicts the data acquired by means of
the Rosenbluth technique according to which $\ge$ and $\gm$) up to several GeV$^2$
approximately follow the dipole form and, hence, $\mugegm \approx 1$.

In this work, we propose a new independent method for measuring the squared Saks form factors.
In this approach, they can be determined separately and independently by direct measurements
of the cross sections without and with spin flip of the initial proton, which should be at
rest and fully polarized in the direction of the motion of the scattered proton. For the case
of the partially polarized initial proton, we propose measuring the ratio of these cross
sections, which makes it possible to determine the ratio of the squared Saks form factors.
This is the aim of our work.

According to Eq. (\ref{Ross}), when the initial and final protons are fully polarized, the ratio of
the cross sections without and with proton spin flip has the extremely simple form
\ba
\frac{d\sigma_{\uparrow\uparrow}}{d\sigma_{\uparrow\downarrow}}
=\frac{\varepsilon}{\tau}\,\frac{G_E^2}{G_M^2}\;.
\label{rat1}
\ea
The simplest way for verifying the correctness of Eqs. (\ref{Ross}) and (\ref{rat1}) is to use the method
for calculating the matrix elements in the diagonal spin basis \cite{GS98} [see Eq. (\ref{Tsquare})], which allows
one to project the spins in the initial and final states of the particles onto one common
direction. The generalization of Eq. (\ref{rat1}) for the most general case of the partially polarized
initial proton is given below [see Eq. (\ref{gen})].

Our proposals are based on the results obtained in \cite{Sikach84}, where it was shown that the matrix
elements of the proton current in the diagonal spin basis that correspond to the transitions
without and with proton spin flip are expressed only in terms of the electric, $G_E$, and magnetic,
$G_M$, Saks form factors, respectively. Note that our terminology for the cross sections
(amplitudes) with and without proton spin flip is not conventional, but has an absolute
physical meaning, because we choose one common direction of the spin projection for the
initial and detected protons; this direction coincides with the direction of the motion
of the scattered proton.

The corresponding experiment on the measurement of the squared Saks form factors in the processes
with and without proton spin flip can be performed as follows. The initial proton at rest and
detected proton should be partially polarized along the direction of the scattered proton or
in the opposite direction. Measuring the corresponding differential cross section, one can
determine the ratio of the squared Saks form factors. The proposed method can be applied to
the elastic muon--proton scattering and implemented in the COMPASS experiment.
The mechanism under consideration is also present in the radiative $ep$ scattering.
In the Bethe--Heitler kinematics, where the leading contribution to the process cross section
comes from two diagrams corresponding to the emission of a photon by an electron, the above
consideration for the elastic $ep$ scattering remains applicable.
The ratio of the squares of the proton electric and magnetic form factors can also be measured
for the process of the photoproduction of lepton pairs on a polarized proton in the
Bethe--Heitler kinematics.
In this work, we consider only the mechanism of the single-photon exchange between the electron
and proton. Our consideration is inapplicable for the two-photon exchange. However,
the contribution of the two-photon mechanism (caused by the interference of the amplitudes
with the exchange by one and two photons) is about 0.5\% of the contribution from the
single-photon mechanism.

\section*{Matrix elements of the proton current in the diagonal spin basis}

In the Born approximation, the matrix element corresponding to the electron--proton elastic scattering,
\ba
e(p_1)+p(p,a) \to e(p_{2}) + p(p',a')
\label{EPEP}
\ea
where $a$ and $a'$ are the polarization 4-vectors of the initial and final protons, has the form
\ba
&& M_{ep\to ep} = \overline{u}(p_{2}) \gamma^{\mu} u(p_{1}) \cdot \overline{u}(p')
\Gamma_{\mu}(q^{2}) u(p) \;  \frac {1} {q^{2}} \; ,\label{5.2}\\
%\ea
%\ba
&& \Gamma_{\mu}(q^{2}) = F_{1} \; \gamma_{\mu} + \frac{F_{2}} {4M}
( \; \hat q \gamma_{\mu} - \gamma_{\mu} \hat q \; ) \; , \; q=p\,'-p \;,
\label {5.4}
\ea
with the mass-shell conditions $p_1^2=p_2^2=m^2$ for electrons and $p^2=p^{'2}=M^2$ for protons.
The matrix elements of the proton current corresponding to the transitions without and with spin flip,
\ba
( J^{\pm\delta ,\delta }_{p} )_{\mu} = \overline{u}^{\pm \delta }(p')
\Gamma_{\mu}(q^{2}) u^{\delta }(p) \; ,
\label {2.3}
\ea
calculated in the diagonal spin basis \cite{Sikach84,GS98} can be expressed in terms of the Saks form factors,
\ba
G_{E} = F_{1} -\tau F_{2} \, , \;
G_{M} = F_{1} + F_{2} \;,
\label {curr}
\ea
where $F_{1}$ and $F_{2}$ are the Dirac and Pauli proton form factors, respectively. The matrix
elements of the proton current given by Eq. (\ref{2.3}) in the diagonal spin basis have the form \cite{Sikach84,GS98}
\ba
( J^{\delta ,\delta }_{p} )_{\mu} &= & 2 G_{E} M ( b_{0} )_{\mu} \, , \label{5.19 non}\\
( J^{-\delta ,\delta }_{p} )_{\mu}& =& - 2 \delta^{} M \sqrt{\tau} G_{M} (
b_{\delta } )_{\mu} \; ,
\label {5.19}
\ea
where
\ba
&& b_{0} = (p+p\,')/\sqrt{(p+p\,')^2} \; , \; b_{3} = q/\sqrt{Q^2} \; ,\nn \\
&&(b_1)_{ \mu} = \varepsilon_{\mu \nu \kappa \sigma}b_0^{\nu}b_3^{\kappa}b_2^{\sigma} \;,
(b_{2})_{\mu} = \varepsilon_{\mu \nu \kappa \sigma}p^{\nu}p^{\,'\kappa}p_1^{\sigma}/\rho , \label {5.9}\\
&& b_{\delta} = b_{1} + i \delta b_{2} \; ,  \delta = \pm 1 \; ,
\; b_{1}^{2} = b_{2}^{2} = b_{3}^{2} = - b_{0}^{2} = - 1 \; .\nn
\ea
Here, $\varepsilon_{\mu\nu\rho\sigma}$ is the Levi-Civita tensor ($\varepsilon_{0123}=-1$),
$\rho$ is determined from the normalization conditions, and the set of unit 4-vectors
$ b_0, b_1, b_2, b_3$ is an orthonormalized basis, $b_{\delta}^{\ast} = b_{1} - i \delta b_{2}$.

Therefore, the matrix elements of the proton current in the diagonal spin basis that correspond
to the transitions without and with proton spin flip given by Eqs. (\ref{5.19 non}) and (\ref{5.19}) are expressed
in terms of the electric, $G_{E}$, and magnetic, $G_{M}$, form factors, respectively (see \cite{Sikach84, GS98}).
In the Breit system of the initial and final protons, which is a particular case of the
diagonal spin basis, Eqs. (\ref{5.19 non}) and (\ref{5.19}) coincide with similar Eq. (4.55) from \cite{RekBook}
and Eqs. (8.55) and (8.56) from \cite{Kallen}.

In the diagonal spin basis \cite{Sikach84,GS98}, the spin 4-vectors $a$ and $a '$ of the protons
with the 4-momenta $p$ and $p\,'$ ($ap = a\,'p\,' = 0 , a ^{2} = a ^{\,'2} = - 1 $),
respectively, lie in the hyperplane formed by the 4-vectors $p$ and $p\,'$:
\ba
\label {DSB}
a = - \frac { (v v') v - v'}  {\sqrt{ ( v v' )^{2} - 1 }} \; , \;
a' =  \frac{ (v v') v' - v } {\sqrt{( v v' )^{2} - 1 }}  \; ,
 v = \frac{p}{M}\,, \; v \,'= \frac{p\,'}{M}\; .
\ea
Spin 4-vectors (\ref{DSB}) obviously remain unchanged under the transformations of the small
Lorentz group common for the particles with the 4-momenta $p$ and $p\,'$. Thus, the
diagonal spin basis provides the description of the spin states of the system of
two particles by means of the spin projections onto one common direction. Note that
this common direction of the projection of the spins of the initial and final protons
in the rest frame of the initial proton is the direction of the motion of the final
proton (see Appendix).

The fundamental fact of the existence of the small Lorentz group common for the particles
with the momenta $p$  and $p\,'$ in the diagonal spin basis gives rise to a number of
remarkable features. First, the particles with the 4-momenta  $p$ (before interaction)
and $p\,'$ (after interaction) in the diagonal spin basis have common spin operators.
This makes it possible to separate the interactions with and without change in the spin states
of the particles involved in the reaction in the covariant form and, thus, to trace the
dynamics of the spin interaction. The spin states of massless particles in the diagonal
spin basis coincide up to sign with helical states \cite{GS98}; in this case, the diagonal spin
basis formalism is equivalent to the CALKUL group method \cite{Berends}.

\section*{Generalization of the diagonal spin basis to the case of partially
polarized protons}

The general expression for the ratio of the cross sections for the $ ep \to ep $
process without and with spin flip for the partially polarized initial and final protons
has the form (see Appendix for details of the derivation):
\ba
\label{gen}
\frac{d \sigma_{\uparrow \uparrow  } }  {d \sigma_{\uparrow \downarrow} }=
\frac{G_E^2(1+\eta)+(\tau/\varepsilon)G_M^2(1-\eta)} {G_E^2(1-\eta)+
(\tau/\varepsilon)G_M^2(1+\eta}\;, \;\eta=\lambda\lambda'.
\ea
where $\lambda$ and $\lambda'$ are the degrees of the polarization of the
initial and final protons in the direction of the motion of the final proton
($\lambda, \lambda'\leq 1$).

For the peripheral processes of the radiative electron proton--scattering and production
of pairs by a photon on a polarized proton at high energies, we can set $\varepsilon=1$.
In this approximation, relation (\ref{gen}) is valid not only for the elastic process
$ ep \to ep $, but also for the $e p \to ep \gamma$
and $\gamma p \to e \bar e p$ -- processes (see below). To separate elastic
events on a proton, it is necessary to measure the spectrum of elastic electrons in the
radiative $e p$ scattering or distribution over the fraction of the energy of the component
of a pair produced in the photoproduction process.

Let us consider the radiative electron--proton scattering
\ba
e(p_1)+p(p,a) \to e(p_2) + p (p',a') + \gamma(k,e)\,,
\label{EPG}
\ea
where $a$ and $a'$ are the polarization 4-vectors of the initial and final protons,
respectively, and $e$ is the polarization 4-vector of the photon such that
$ap=a'p'=ek=0$. In the peripheral (Bethe--Heitler) kinematics specified by the relations
\ba
s_e=2p_1p>>Q^2=-q^2=-(p-p')^2\sim M^2,
\ea
the matrix element of process (\ref{EPG}) has the factorized form (this is easily shown using the
Gribov identity \cite{baier81}):
\ba
M_e=2s_e\frac{(4\pi\alpha)^{3/2}}{q^2}N_e N^e_p\,,
\ea
where
\ba
&& N^e_p=\frac{1}{s_e}\bar{u}(p',a')\hat{p}_1
\left(F_1(q^2)-\frac{1}{2M}\,F_2(q^2)\hat{q}\right)u(p,a),\nn \\
&&N_e=\frac{1}{s_e}\bar{u}(p_2)\left(\hat{p}\frac{\hat{p}_2-\hat{q}+m}{d_2}\hat{e}+
\hat{e}\frac{\hat{p}_2-\hat{q}+m}{d_2}\hat{p}\right)u(p_1).\nn
\ea
Similar expressions can be written for the matrix element for the production of pairs
by a photon on a polarized proton:
\ba
\gamma(k,e)+p(p,a) \to e^+(p_+) + e^-(p_-)+p(p',a')\,.
\label{GPP}
\ea
In the Bethe--Heitler kinematics,
\ba
s_\gamma =2kp >> Q^2\sim M^2, \nn
\ea
the matrix element of process (\ref{GPP}) has the form
\ba
M_\gamma=2s_\gamma\frac{(4\pi\alpha)^{3/2}}{q^2}N_\gamma N^\gamma_p\,,
\ea
where
\ba
&&N^{\gamma}_p=\frac{1}{s_\gamma}\bar{u}(p',a')\hat{k}
\left(F_1(q^2)-\frac{1}{2M}F_2(q^2)\hat{q}\right)u(p,a), \nn\\
&&N_\gamma=\frac{1}{s_\gamma}\bar{u}(p_-)\left(\hat{p}\frac{\hat{k}-\hat{p}_++m}{d_+}\hat{e}+
\hat{e}\frac{\hat{p}_--\hat{k}+m}{d_-}\hat{p}\right)v(p_+)\,.\nn
\ea
Calculating the squared absolute values of the matrix elements of the proton current,
$|N^e_p|^2$ and $|N^\gamma_p|^2$, for processes $ep\to ep\gamma$ (\ref{EPG})
and $\gamma p \to e \bar e p$ (\ref{GPP}) without and with spin flip,
we arrive at the same expressions
\ba
&&  |N^e_p(a,\pm a')|^2=|N^\gamma_p(a,\pm a')|^2=4G_\pm \,,  \nn \\
&&G_\pm=\frac{1}{2(1+\tau)}\left(G_E^2(1 \pm \eta)+\tau G_M^2 (1\mp \eta)\right) \nn.
\ea
Averaging and summing the expression for $|N_e|^2$  over the spin states of the electrons and photon, we obtain
\ba
&&\sum |N_e|^2=4D_e,  \nn \\
&&D_e=x(1-x)^2\left(\frac{Q^2(1+x^2)}{d_1d_2}-2m^2x(\frac{1}{d_1}
-\frac{1}{d_2})^2\right)\,,\nn
\ea
where
\ba
&&d_1=m^2(1-x)^2+\vec{p}^{~2}_2, \nn\\
&&d_2=m^2(1-x)^2+(\vec{p}_2+\vec{q}(1-x))^2,\nn
\ea
and $x$ is the energy fraction carrying by the scattered electron,
$\vec{p}_2$ is the momentum component transverse to the electron beam, and $\vec{q}$ is the transverse
momentum component transferred to the proton.

Finally, averaging and summing the expression for $|N_\gamma|^2$ over the spin states
of the components of a pair and the photon for $\gamma p \to e \bar e p$ process (\ref{GPP}), we obtain
\ba
&&\sum |N_\gamma|^2=4D_\gamma \,, \nonumber \\
&&D_\gamma=x_+x_-\left(\frac{Q^2(x_+^2+x_-^2)}{d_+d_-}+2m^2x_+x_-(\frac{1}{d_+}-\frac{1}{d_-})^2\right),\nn
\ea
where $x_-$ and $x_+$ are the energy fractions carried by the electron and positron
($x_+ + x_-=1$), respectively; $d_\pm=\vec{p}_\pm^{~2}+m^2$; and $\vec{p}_-$ and $\vec{p}_+$
are the transverse momenta of the components of the pair ($\vec{p}_- +\vec{p}_+=\vec{q}$).

Relation (\ref{gen}) is valid for processes $ep\to ep\gamma$ (\ref{EPG})
and $\gamma p \to e \bar {e} p $ (\ref{GPP}), because the differential cross
sections for these processes in the Bethe--Heitler kinematics have the form
\ba
\label{epepg aa}
d\sigma^{ep\to ep\gamma}(a,\pm a')=\frac{2\alpha^3}{\pi^2(Q^2)^2}
D_e G_\pm \frac{d^2q d^2p_2 dx}{x(1-x)}\;,
\ea
\ba
d\sigma^{\gamma p\to e\bar{e} p}(a,\pm a')=\frac{2\alpha^3}{\pi^2(Q^2)^2}
D_\gamma G_\pm \frac{d^2q d^2p_- dx_-}{x_-x_+}\;.
\ea
The integration of differential cross section (\ref{epepg aa}) with respect to the
transverse momentum of the final electron yields
\ba
\frac{d\sigma^{e p\to e p \gamma}}{dQ^2dx}(a,\pm a')=\frac{2\alpha^3}{(Q^2)^2}G_\pm
\times[[\tau_1(1+x^2)+x]R(\tau_1)-x]\;,
\ea
where $\tau_1=Q^2/m^2$ and
\ba
R(z)=\frac{1}{\sqrt{z(1+z)}}\ln(\sqrt{1+z}+\sqrt{z})\;.
\ea
A similar expression for the cross section for the photoproduction of pairs on the proton has the form
\ba
\frac{d\sigma^{\gamma p\to e\bar{e} p}}{dQ^2dx_-}(a,\pm a')=
\frac{2\alpha^3}{(Q^2)^2}G_\pm
\times [[\tau_1(x_+^2+x_-^2)-x_+x_-]R(\tau_1)+x_+x_-]\;. %\nn
\ea
The details of the $ep \to ep$ kinematics in the rest frame of the
initial proton are discussed in the Appendix.

\section*{Conclusion}

Direct relations between the ratio of the squared proton form factors and the ratio of the
cross sections for the processes without and with proton spin flip have been obtained in
the diagonal spin basis approach. Similar relations for the radiative electron--proton
scattering and photoproduction of pairs on a polarized proton exist in Bethe--Heitler
kinematics. We emphasize that, in view of Eqs. (\ref{5.19 non}) and (\ref{5.19}) for the matrix elements
of the proton current in the diagonal spin basis, the differential cross section
for the elastic scattering of the electron on the fully polarized proton without proton
spin flip is expressed only in terms of the square of the electric form factor
$\ge$, whereas the differential cross section for the process with proton spin
flip is expressed only in terms of the square of the magnetic form factor $\gm$
for any square of the momentum transferred to the proton. For this reason, the
experiment on the measurement of the cross sections for the process on the fully
polarized proton without and with proton spin flip would have a number of features
as compared to the Rosenbluth method, which is applicable only for low $Q^2$ values,
and the method of the polarization transfer from the initial electron to the final
proton \cite{Rekalo}. To carry out such an experiment, the initial proton at rest should be
fully polarized in the direction of the motion of the scattered proton. Such an
experiment would provide new independent data on the behavior of the form factors
$G_E^2(Q^2)$ and $G_M^2(Q^2)$ and assist to resolve the contradiction appearing
after the know experiment of the JLab collaboration \cite{PRL}.
%\vspace{-0.5cm}

\hspace*{7cm}
\begin{flushleft}
\hspace*{15cm}{\bf Appendix}
\end{flushleft}
\vspace{-0.5cm}
\section*{Kinematics of the Elastic Process in the Laboratory Reference Frame}
%\label{RecoilProtonDetectionKinematics}
In the laboratory reference frame where the initial proton is at rest,
$p=M(1,0,0,0)$, the polarization 4-vectors $a$ and $a'$ of the initial
and final protons, respectively, in the diagonal spin basis given by Eqs. (\ref{DSB})  and the
4-momentum of the final proton $p\,'$ have the form
\ba
&&~~~a=(0,\vec{a})=(0, \vec{n}), \,a'=\frac{1}{M}(p, E' \vec{n})\,,\nn\\
%\ea
%\ba
&&p\,'=(E', p\vec{n}), E'=M(1+2\tau), p=2M\sqrt{\tau(1+\tau)},\nn
\ea
where $\vec{n}$ is the unit vector along the direction of the motion of the final proton.
Thus, the spin 4-vectors of the diagonal spin basis for the initial and final protons
are defined so that the axes of the spin projections of the protons in the laboratory
reference frame under consideration coincide and are directed along the momentum of
the final proton: $\vec{a}=\vec{n}$. In this case, the spin state of the final proton is helical.
The 4-momentum of the initial electron has the form $p_1=E_e(1,0,0,1)$.

Calculating the convolution $I=L^{\mu\nu}P_{\mu\nu}$ of the lepton,
\ba
L^{\mu\nu}&=&p_1^\mu p_2^{\nu}+p_1^\nu p_2^{\mu}+q^2g^{\mu\nu}/2 \,,\nn
\ea
and proton,
\ba
P_{\mu\nu}=\frac{1}{4}Tr (\hat{p}'+M)(1-\gamma_5\hat{a}')\gamma_\mu \left(F_1-\hat{q}\frac{1}{2M}F_2\right)
(\hat{p}+M)(1-\gamma_5\hat{a})\gamma_\nu\left(F_1+\hat{q}\frac{1}{2M}F_2\right)\,,\nn
\ea
tensors and using the kinematic relations
\ba
&&(aa')=-1-2\tau,\,(ap') (a'p)=-4M^2\tau(1+\tau),\, \nn \\
&&(ap') (ap_1)=\tau[2M^2(1+2\tau)-s]\,, \nn \\
&& (ap_1) (a'p)= -2M(E+M)\tau,\;\nn \\
&& (ap_1) (a'p_1)=\frac{\tau}{1+\tau}(E+M)[-E+M(1+2\tau)]\,,\nn
\ea
we arrive at result (\ref{rat1}). To generalize the diagonal spin basis to the case of the
partially polarized states of the initial and final protons and to derive the
corresponding general expressions for the ratio of the cross sections without and
with proton spin flip, it is necessary to change $a \to \lambda a$
and $a\,'\to \lambda\,'a\,'$, $\lambda, \lambda'\leq 1$.
As a result, we obtain Eq. (\ref{gen}).

The energy and 3-momentum of the recoil proton in the laboratory reference frame are
expressed in terms of its scattering angle with respect to the direction of the motion
of the initial electron $\theta_p$ (see \cite{VK72}) as
\ba
\frac{E'}{M}=\frac{1+\cos^2\theta_p}{\sin^2\theta_p}\,,\;
\frac{p}{M}=\frac{2\cos\theta_p}{\sin^2\theta_p}\;.\nn
\ea

The use of the matrix elements of the proton current given by Eqs. (\ref{5.19 non}) and (\ref{5.19}) reduces
the calculation of the probability of the $ep \to ep$ process to the
calculation of the trivial trace
\ba
\label{Tsquare}
\mid T \mid^2=\frac{4M^2}{q^4}\, \frac{1}{8}\,\sum_{\delta} Tr \left( G^2_E(\hat p_2 +m)\hat b_0(\hat p_1 +m)\hat b_0+
 \tau G^2_M(\hat p_2 +m)\hat b_{\delta}(\hat p_1 +m)\hat b^*_{\delta} \right)\;.\nn
\ea
As a result, we arrive at the following expression coinciding with the result presented in \cite{Rosen}:
\ba
\label{6.1}
&& ~~~~~~~~d \sigma = \frac{ \alpha^{2} d o}{4w^2} \frac{1}{1+\tau} \, ( \; G_{E}^{2} \; Y_{I}
+ \tau \; G_{M}^{2} \; Y_{II} \; )\, \frac{1}{q^{4}}\; , \\
&&Y_I=(p_+ q_+)^2+q_+^2q^2,\; \quad Y_{II}=(p_+ q_+)^2-q_+^2(q^2+4 m^2)\;,   \nn\\
&& ~~~~~~~~~~~\quad  p_+=p_1+p_2\;,   \;  q_+=p +p\,'\,,\nn
\ea
where $d o$ is the solid angle element and $w$ is the total energy in the center-of-mass system.

The ratio of the cross sections for the processes without and with proton spin flip is given by the expression
\ba
\label{genn}
\frac{d \sigma_{\uparrow \uparrow  } }  {d \sigma_{\uparrow \downarrow} }=
\frac{Y_I} {\tau Y_{II}} \,\frac{G_E^2} {G_M^2} \;.
\ea
In the rest system of the initial proton, neglecting the electron mass, we obtain Eqs. (\ref{Ross}) and (\ref{rat1}).

% --------------------------------------------------------------------
\section*{Acknowledgements}
% --------------------------------------------------------------------
We are grateful to Prof. E. Tomasi-Gustafsson for interest in this work and critical
comments. This work was supported by INTAS (grant no. 05-1000008-8328) and Belarusian
State Foundation for Basic Research (project no. F07D-005). M.V.G. is grateful to the
Bogoliubov Laboratory of Theoretical Physics, Joint Institute for Nuclear Research,
for hospitality and possibility of performing joint investigations at the initial stage of this work.

\end{document}